\documentclass[aps,pra,showpacs,twocolumn,superscriptaddress]{revtex4}
\usepackage{amsfonts,amssymb,amsmath}
\usepackage{mathtools}
\usepackage[]{graphics,graphicx,epsfig}
\usepackage{amsthm,multirow}
\begin{document}

\title{Macroscopic limit of nonclassical correlations}

\author{Pawe{\l} \surname{Kurzy\'nski}}
\email{cqtpkk@nus.edu.sg}
\affiliation{Centre for Quantum Technologies, National University of Singapore, 3 Science Drive 2, 117543 Singapore, Singapore}
\affiliation{Faculty of Physics, Adam Mickiewicz University, Umultowska 85, 61-614 Pozna\'{n}, Poland}

\author{Dagomir \surname{Kaszlikowski}}
\email{phykd@nus.edu.sg}
\affiliation{Centre for Quantum Technologies, National University of Singapore, 3 Science Drive 2, 117543 Singapore, Singapore}
\affiliation{Department of Physics, National University of Singapore, 2 Science Drive 3, 117542 Singapore, Singapore}

\date{\today}


\begin{abstract}
We consider macroscopic correlations in a bipartite system consisting of $2N$ particles described by a generalised probabilistic theory. In particular, we discuss a case of $N$ PR-boxes shared between two parties. We characterise macroscopic measurements as collective measurements of the same property on all the boxes in the same region. Such measurements are assumed to reveal only the average value of the measured collective property. We show that for two measurements per observer and $N \geq 2$ there always exist a joint probability distribution explaining all observable data and therefore the system admits local hidden variables. Next, we generalise this result to include measurement of fluctuations and additional measurement settings. Finally, we discuss our result in the context of previous works in which contradictory results were presented.

\end{abstract}

\maketitle


\section{Introduction}

Macroscopic systems consist of many particles. It is extremely difficult, if possible at all, to register individual properties of each microscopic constituent. Therefore our observations are limited only to collective properties of multi-particle ensembles and it is an interesting question whether microscopic effects, like strong non-classical correlations, can be still observed at the macroscopic level. 


The above problem has been considered in literature before, see for example \cite{Navascues,macrobell,Bancal,l1,l2,l3,l4,l5,l6,l7,Rohrlich,Gisin}. In particular, it was shown that in the macroscopic limit, where measurements are limited to reveal only average values \cite{macrobell} and its fluctuations \cite{Navascues}, collective quantum correlations become local, although individual pairs of particles can manifest nonlocality. It is still an open question whether similar \emph{macroscopic locality} \cite{Navascues} also holds for stronger than quantum correlations \cite{PRbox}. This problem has been recently addressed by Rohrlich \cite{Rohrlich} and Gisin \cite{Gisin} who argued that such correlations are unphysical since they allow for signalling in the macroscopic limit. We show the contrary: there exist a local and realistic macroscopic limit of stronger than quantum correlations without signalling.

We generalise our previous result \cite{macrobell}, which applies only to quantum systems, and show that average values of macroscopic measurements on a bipartite system of $2N$ particles, whose correlations are described by generalised probabilistic theories, admit a local and realistic description. In particular, we show that for two measurements per observer for $N \geq 2$ there exists a joint probability distribution (JPD) for all observables \cite{Fine} reproducing measurable average values. Next, we make an extension to more measurement settings and to macroscopic measurements that reveal information about statistical fluctuations. Finally, we compare our results to that of Rorlich and Gisin \cite{Rohrlich,Gisin}. 


\section{Macroscopic measurements}

Let us define macroscopic measurement. Following previous works \cite{Navascues,macrobell,Bancal,l1,l2,l3,l4,l5,l6,l7,Rohrlich,Gisin}, it is a collective measurement of the same property on all $N$ particles, i.e.,
\begin{equation}\label{macro}
A=a^{(1)} + a^{(2)} + \ldots + a^{(N)},
\end{equation}
where $a^{(i)}$ is a microscopic observable on the i-th particle. Here we assume that microscopic observables are binary with $\pm 1$ outcomes. There are $2^{N}$ microscopic states of such a system and they give $N+1$ different outcomes of a macroscopic measurement: $N, N-2, \ldots, -N$. Although $N+1 \ll 2^N$ we assume that it is still impossible to trace all the macroscopic outcomes in realistic measurements and that we are limited to measurements of the $A$'s average value $\langle A \rangle$. Measurements of $A$'s fluctuations will be discussed later.

To justify our definition of macroscopic measurement we use an analogy to macroscopic measurement of a light beam's polarisation. Although the beam is composed of a large number of photons, the macroscopic measurement of its polarisation (say horizontal and vertical components) reveals only intensities $I_H$ and $I_V$. In such measurements one can also observe fluctuations around the mean value but these fluctuations are mainly caused by the internal noise of measurement devices rather than by actual fluctuations of the observed property. 

An interesting problem turns up when a macroscopic system is divided into two parts. One can measure different property on each part, say $A$ and $B$ (defined in a similar way), and ask if the correlations between these properties admit local hidden variables. It is a non-trivial question since we know that the correlations between the microscopic constituents in these divisions may not admit such a description. 

There is one important obstacle in an experimental implementations of this scenario - there is no known method to measure $\langle AB \rangle$ without invoking nonlocal measurements. In order to measure $\langle AB \rangle$ one uses a probe that interacts with both partitions, say first with $A$ and later with $B$. The final measurement of the probe reveals $\langle AB \rangle$ (a possible implementation technique would be similar to the one considered in \cite{Polzik}). In this case there is an obvious locality loophole, since the probe can transfer some information from $A$ to $B$. On the other hand, if one used two separate probes, a different probe for each partition, the first probe would reveal $\langle A \rangle$ and the second $\langle B \rangle$ but not $\langle AB \rangle$. 

Because of the locality loophole the above scenario would rather test the contextuality \cite{KS} of the system than its lack of a local and realistic description. However, even if local measurements of $\langle AB \rangle$ were possible, we show that there exist a JPD reproducing average values of macroscopic measurements. Thus, even if microscopic correlations are not local and realistic, the nonclassical effects resulting from these correlations are unobservable on a macroscopic level - they can be reproduced by a system admitting local and realistic/non-contextual description.


\section{Main Result}


\subsection{Macroscopic correlations}

Let us consider $2N$ particles shared by Alice and Bob. In general, the correlations between these particles may not be classical and not even quantum, but can be described by general probabilistic theories \cite{PRbox}. Each particle can be measured in two different settings $a_0^{(k)}$ and $a_1^{(k)}$ for the Alice's k-th particle and $b_0^{(l)}$ and $b_1^{(l)}$ for the Bob's l-th particle. We assume that all measurements are binary with $\pm1$ outcomes.

We can calculate the following probabilities
\begin{equation}\label{pm}
p(a_{i_1}^{(1)}=x_{i_1}^{(1)},\ldots,a_{i_N}^{(N)}=x_{i_N}^{(N)};b_{j_1}^{(1)}=y_{j_1}^{(1)},\ldots,b_{j_N}^{(N)}=y_{j_N}^{(N)}),
\end{equation}
where $i_k,j_l = 0,1$ label measurement setting ($k,l=1,\ldots,N$) and $x_{i_k}^{(k)},y_{j_l}^{(l)}=\pm 1$ are corresponding outcomes. We will write (\ref{pm}) as
\begin{equation}\label{pmicro}
p(x_{i_1}^{(1)},\ldots,x_{i_N}^{(N)};y_{j_1}^{(1)},\ldots,y_{j_N}^{(N)}).
\end{equation}

We also assume that these probabilities are non-signalling, i.e., the same marginal probability obtained from two different global probabilities are equal. For example 
\begin{eqnarray}
& &\sum_{x_0^{(1)}=\pm 1} p(x_{0}^{(1)},\ldots,x_{i_N}^{(N)};y_{j_1}^{(1)},\ldots,y_{j_N}^{(N)}) = \nonumber \\
& &\sum_{x_1^{(1)}=\pm 1} p(x_{1}^{(1)},\ldots,x_{i_N}^{(N)};y_{j_1}^{(1)},\ldots,y_{j_N}^{(N)}).
\end{eqnarray}

However, as macroscopic measurements (\ref{macro}) do not address each particle individually, we are limited to a subset of probability distributions in which all settings in a partition are the same, i.e., $i_1=\ldots=i_N=i$ and $j_1=\ldots=j_N=j$. Thus, we are interested in the following probabilities
\begin{equation}\label{pmicro2}
p(x_{i}^{(1)},\ldots,x_{i}^{(N)};y_{j}^{(1)},\ldots,y_{j}^{(N)}).
\end{equation}

In our scenario there are four macroscopic measurements 
\begin{equation}\label{macro2}
A_i = \sum_{k=1}^N a_i^{(k)},~~B_j = \sum_{l=1}^N b_j^{(l)},
\end{equation}
where $i,j=0,1$. The measurement of $\langle A_i B_j\rangle$ yields
\begin{equation}
\langle A_i B_j \rangle = \sum_{k,l=1}^N \langle a_i^{(k)}b_j^{(l)} \rangle,
\end{equation}
where
\begin{eqnarray}
\langle a_i^{(k)}b_j^{(l)} \rangle &=& p(x_i^{(k)}=+1;y_j^{(l)}=+1) \nonumber \\ 
&+& p(x_i^{(k)}=-1;y_j^{(l)}=-1)  \nonumber \\ 
&-& p(x_i^{(k)}=+1;y_j^{(l)}=-1)  \nonumber \\ 
&-& p(x_i^{(k)}=-1;y_j^{(l)}=+1), \label{average}
\end{eqnarray}
and $p(x_i^{(k)};y_j^{(l)})$ is a marginal of (\ref{pmicro2}). 

\begin{figure}[t]
    \begin{center}
    	\includegraphics[width=1.0\columnwidth,trim=4 4 4 4,clip]{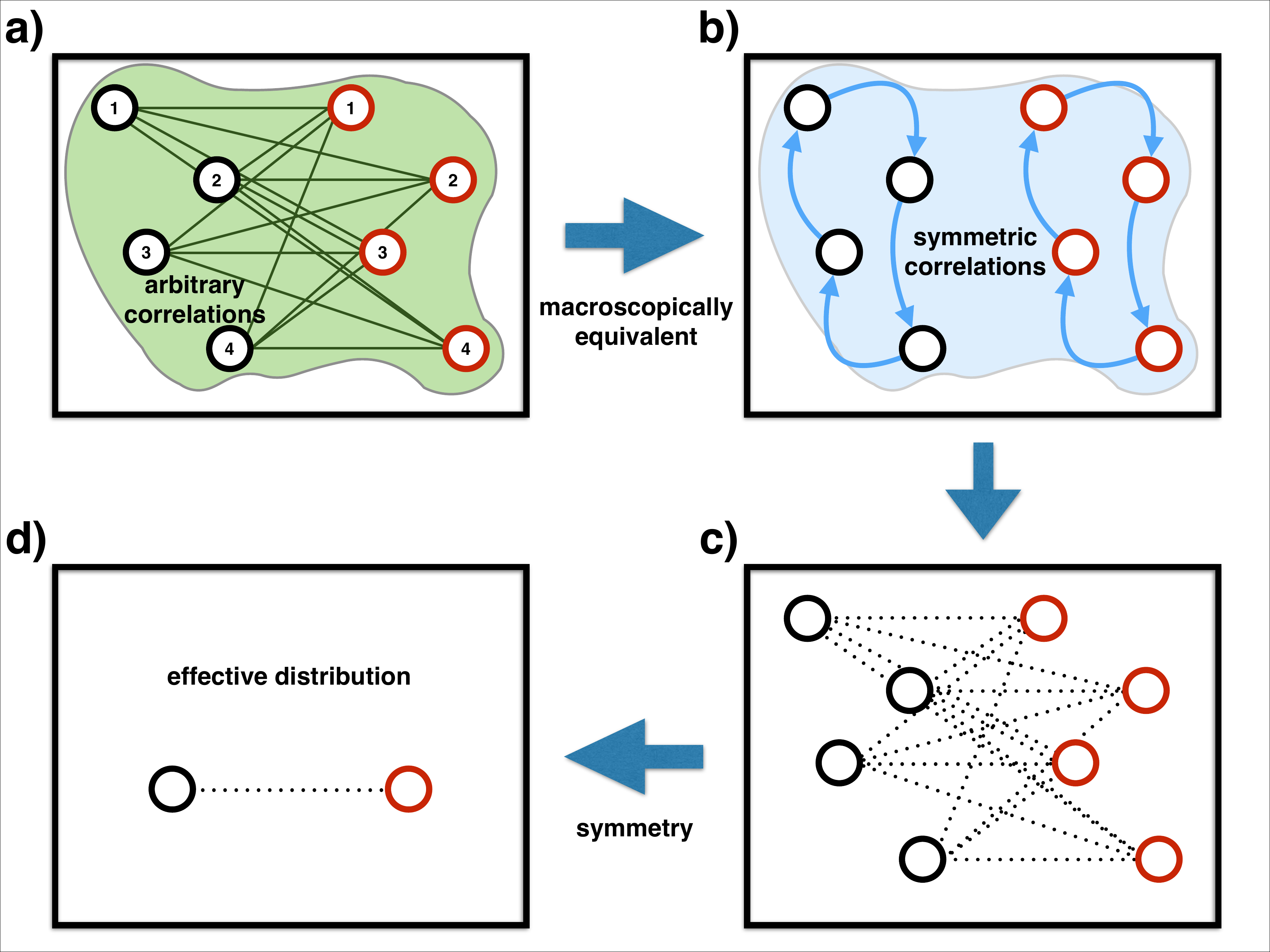}
        \caption{Effective description of macroscopic measurement of $\langle AB \rangle$. a) The original distribution describing arbitrary correlations. b) Macroscopic measurements cannot distinguish between the original distribution and its symmetrised version. c) Symmetric correlations are the same for every pair. d) Effective simulation with a single pair.}
        \label{fig1}
    \end{center}
\end{figure}

An interesting feature of macroscopic measurements (\ref{macro2}) is that the macroscopic outcome stays the same under the particle exchange. In particular, if we consider all possible permutations of particles in each region then microscopic correlations between every pair will be given by an effective correlation
\begin{equation}
\langle a_i b_j \rangle_{eff} \equiv \frac{1}{N^2}\sum_{k,l=1}^N \langle a_i^{(k)}b_j^{(l)} \rangle,
\end{equation}
resulting from an effective probability
\begin{equation}\label{peff}
p(x_i;y_j)_{eff} \equiv \frac{1}{N^2}\sum_{k,l=1}^N p(x_i^{(k)}=x_i;y_j^{(l)}=y_j).
\end{equation}
Thus
\begin{equation}
\langle A_i B_j \rangle = N^2 \langle a_i b_j \rangle_{eff}.
\end{equation}
It is clear that $\langle A_i B_j \rangle$ is simulated by effective correlations between a single pair of particles, see Fig. \ref{fig1}.


\subsection{Joint probability distribution for correlations}

Although individual particle addressing is not allowed in the macroscopic limit, marginal probabilities of (\ref{pmicro}) exist
\begin{equation}
p(x_0^{(k)},x_1^{(l)};y_0^{(m)},y_1^{(n)}).
\end{equation}
We can use them to construct a symmetric probability distribution
\begin{eqnarray}
& & p(x_0,x_1;y_0,y_1)_{sym} = \frac{1}{N^2 (N-1)^2} \times \label{psymm} \\
& & \sum_{\substack{k,l|k\neq l \\ m,n|m\neq n}} p(x_0^{(k)},x_1^{(l)};y_0^{(m)},y_1^{(n)}). \nonumber
\end{eqnarray}
This distribution naturally arises under the symmetrisation over all particles in each region.

However, it is straightforward to show that 
\begin{eqnarray}
p(x_0;y_0)_{eff} &=& \sum_{x_1,y_1} p(x_0,x_1;y_0,y_1)_{sym}, \nonumber \\
p(x_0;y_1)_{eff} &=& \sum_{x_1,y_0} p(x_0,x_1;y_0,y_1)_{sym}, \nonumber \\
p(x_1;y_0)_{eff} &=& \sum_{x_0,y_1} p(x_0,x_1;y_0,y_1)_{sym}, \nonumber \\
p(x_1;y_1)_{eff} &=& \sum_{x_0,y_0} p(x_0,x_1;y_0,y_1)_{sym}.
\end{eqnarray}
Therefore, (\ref{psymm}) is a joint probability distribution that reconstructs measurable average values and hence these values admit a local and realistic/non-contextual description \cite{Fine}.


\subsection{Macroscopic fluctuations}

We now deal with macroscopic fluctuations around the average values as in principle they can be measured too. The probability distribution (\ref{psymm}) recovers macroscopic correlations $\langle A_i B_j \rangle$ and local averages $\langle A_i \rangle$ and $\langle B_j \rangle$ but not their fluctuations 
\begin{equation}
\Delta( A_i) = \sqrt{\langle A_i^2 \rangle - \langle A_i \rangle^2}
\end{equation}
and
\begin{equation}
\Delta (A_i B_j) = \sqrt{\langle (A_i B_j)^2 \rangle - \langle A_i B_j \rangle^2}
\end{equation}
Therefore, we need to look for a way to reconstruct $\langle A_i^2 \rangle$ and $\langle (A_i B_j)^2 \rangle$.

First, note that
\begin{equation}\label{localfluctuations}
\langle A_i ^2 \rangle = \sum_{k,l=1}^{N} \langle a_i^{(k)}a_i^{(l)} \rangle = N  +  \sum_{k,l|k\neq l} \langle a_i^{(k)}a_i^{(l)} \rangle 
\end{equation}
and
\begin{eqnarray}
\langle (A_i B_j)^2 \rangle &=& \sum_{k,l,m,n=1}^{N} \langle a_i^{(k)}a_i^{(l)}b_j^{(m)}b_j^{(n)} \rangle = N^2  \nonumber \\
&+&  N\sum_{k,l|k\neq l} \langle a_i^{(k)}a_i^{(l)} \rangle + N\sum_{m,n|m\neq n} \langle b_j^{(m)}b_j^{(n)} \rangle \nonumber \\
&+&  \sum_{\substack{k,l|k\neq l \\ m,n|m\neq n}} \langle a_i^{(k)}a_i^{(l)}b_j^{(m)}b_j^{(n)} \rangle \label{fluctuations}
\end{eqnarray}
where the 4-point correlations are given by the marginals of (\ref{pmicro2}), i.e., $p(x_i^{(k)},x_i^{(l)};y_j^{(m)},y_j^{(n)})$. The bipartite correlations $\langle a_i^{(k)}a_i^{(l)} \rangle$ and $\langle b_j^{(m)}b_j^{(n)} \rangle$ are given by $p(x_i^{(k)},x_i^{(l)})$ and $p(y_j^{(m)},y_j^{(n)})$ respectively. 

As before, the macroscopic measurements cannot distinguish between the original probability distribution and the effective one obtained via symmetrisation
\begin{eqnarray}
& & p(x_i,x'_i;y_j,y'_j)_{eff} =  \frac{1}{N^2(N-1)^2}\times  \label{4eff} \\
& & \sum_{\substack{k,l|k\neq l \\ m,n|m\neq n}} p(x_i^{(k)},{x'}_i^{(l)};y_j^{(m)},{y'}_j^{(n)}). \nonumber
\end{eqnarray}
Primes are used to denote two different outcomes on two local observables. Although we distinguish between these outcomes, the effective probabilities obey $p(x_i,x'_i;y_j,y'_j)_{eff}=p(x'_i,x_i;y_j,y'_j)_{eff}=p(x_i,x'_i;y'_j,y_j)_{eff}=p(x'_i,x_i;y'_j,y_j)_{eff}$.

\begin{figure}[t]
    \begin{center}
    	\includegraphics[width=1.0\columnwidth,trim=4 4 4 4,clip]{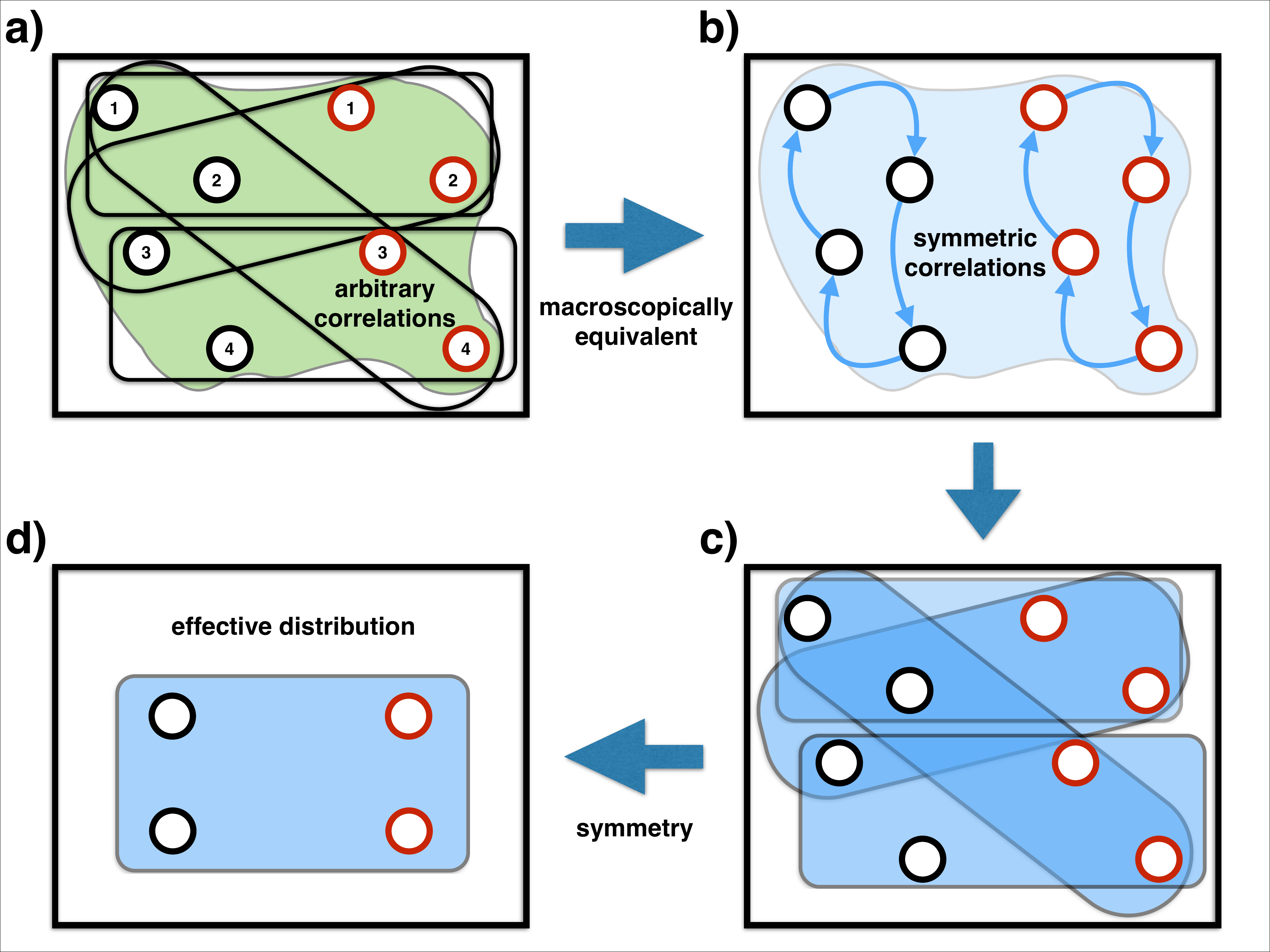}
        \caption{Scheme representing an effective description of a macroscopic measurement of $\langle (AB)^2 \rangle$. a) The original distribution describing arbitrary correlations. b) Macroscopic measurements cannot distinguish between the original distribution and its symmetrised version. c) Symmetric correlations are the same for every two pairs. d) Effective simulation with two pairs.}
        \label{fig2}
    \end{center}
\end{figure}

As a result $\langle (A_i B_j)^2 \rangle$ is effectively simulated by two pairs of particles admitting the distribution $p(x_i,x'_i;y_j,y'_j)_{eff}$,  see Fig. \ref{fig2}. This distribution yields $\langle a_i a'_i b_j b'_j \rangle_{eff}$, $\langle a_i a'_i\rangle_{eff}$, $\langle b_j b'_j \rangle_{eff}$ and allows us to express (\ref{localfluctuations}) and (\ref{fluctuations}) as 
\begin{equation}\label{fl1}
\langle A_i ^2 \rangle = N(1 + (N-1)\langle a_i a'_i\rangle_{eff}) 
\end{equation}
and
\begin{eqnarray}
& &\langle (A_i B_j)^2 \rangle = N^2(N-1) \times \label{fl2} \\ 
& &\left(\frac{1}{N-1}+ \langle a_i a'_i\rangle_{eff} + \langle b_j b'_j \rangle_{eff} + (N-1) \langle a_i a'_i b_j b'_j \rangle_{eff}\right). \nonumber
\end{eqnarray}


Finally, we write a symmetric probability distribution
\begin{eqnarray}
& & p(x_0,x'_0,x_1,x'_1;y_0,y'_0,y_1,y'_1)_{sym} = \frac{((N-4)!)^2}{(N!)^2} \times \label{psym} \nonumber \\
\tilde{\sum} & & p(x_0^{(k)},{x'}_0^{(k')},x_1^{(l)},{x'}_1^{(l')};y_0^{(m)},{y'}_0^{(m')},y_1^{(n)},{y'}_1^{(n')}) \label{psym2} 
\end{eqnarray}
where $\tilde{\sum}$ denotes the sum over $k\neq l \neq k' \neq l'$ and $m\neq n \neq m' \neq n'$. The probabilities $p(x_i,x'_i;y_j,y'_j)_{eff}$ are marginals of (\ref{psym2}). Moreover, the probabilities (\ref{peff}) are also marginals of (\ref{psym2}). This confirms that (\ref{psym2}) is a JPD that reproduces average values and their fluctuations.

At this point we need to comment on an important property of our approach that differs from the standard one \cite{Fine}. For average values we showed that the original system can be effectively simulated by a single pair of particles whose correlations are described by the JPD (\ref{psymm}). To obtain the original average value from the simulated one we only need to rescale the simulated outcome by the factor of $N^2$. Therefore, the average values stem directly from the JPD. 

On the other hand, to simulate fluctuations one needs two effective pairs described by the JPD (\ref{psym2}). However, the original fluctuations cannot be obtained by a simple rescaling of fluctuations of the simulating system. The original fluctuations do not stem directly from the JPD, but are its functions (\ref{fl1}) and (\ref{fl2}). From the practical point of view the local and realistic/non-contextual system of four particles, together with two local devices computing functions of the JPD, can be effectively used to simulate averages and fluctuations of macroscopic measurements on a bipartite system consisting of many nonclassically correlated particles.


\subsection{Higher moments and more measurement setups}

The above results can be generalised to explain the behaviour of higher moments $\langle (A_i B_j)^k \rangle$ and scenarios with more than two measurement setups per observer. Also, note that the above discussion does not depend on the number of measurement outcomes in a given setup.

Without going into details, let us note that our method relies on the symmetrisation of the probability distribution and on choosing a proper subset of particles on which we measure all the observables. For example, for $k=3$ one needs to measure correlations of the form $\langle a_i^{(k)}a_i^{(l)}a_i^{(m)}b_j^{(n)}b_j^{(o)}b_j^{(p)}\rangle$. This means that in case of two measurement setups per observer one needs $3+3+3+3=12$ particles to obtain  the corresponding JPD. 

In general, in order to obtain a JPD simulating the $k$-th moment in case of $s_A$ and $s_B$ measurement setups one needs $k (s_A + s_B)$ particles. This shows that in order to look for macroscopic nonclassicality one needs to make very accurate measurements of fluctuations to reveal higher moments $k (s_A + s_B) > N$. Indeed, if one could measure sufficiently high moments, one would gain enough insight into a macroscopic probability distribution $p(A_i=X_i;B_j=Y_j)$, where $X_i,Y_j=N,N-2,\ldots,-N$, and would be able to apply methods of Ref. \cite{Bancal} to observe violation of local realism. This seems to be unrealistic in a macroscopic limit with Avogadro number $N \sim 10^{23}$ of particles. 


\section{Example -- N pairs of PR-boxes}

Here we show how to apply our result to a specific scenario. Let us consider a bipartite system consisting of N pairs of PR-boxes. PR-box (Popescu-Rohrlich-box) is a hypothetic bipartite system manifesting stronger than quantum correlations without signalling \cite{PRbox}. 


\subsection{Correlations}

Each PR-box pair is described by correlations
\begin{equation}
\langle a_i b_j \rangle_{PR} = (-1)^{i\cdot j},
\end{equation}
and the corresponding probabilities
\begin{equation}
p_{PR}(x_i;y_j) = \frac{|x_i+(-1)^{i\cdot j}y_j|}{4}.
\end{equation}
Local marginal probabilities are maximally random $p_{PR}(x_i)=p_{PR}(y_j)=\frac{1}{2}$, hence $\langle a_i \rangle=\langle b_j \rangle=0$. In the case of $N$ pairs each PR-box pair is independent, i.e., probabilities factorise. For example, for any two pairs $k$ and $l$ we have 
\begin{equation}
p(x^{(k)}_i,x^{(l)}_j;y^{(k)}_m,y^{(l)}_n) = p_{PR}(x_i;y_m)p_{PR}(x_j;y_n).
\end{equation}
The above implies that there are no correlations between boxes corresponding to different pairs since
\begin{equation}
p(x^{(k)}_i;y^{(l)}_n) = p_{PR}(x_i)p_{PR}(y_n)=\frac{1}{4},
\end{equation}
therefore
\begin{equation}
\langle a^{(k)}_ib^{(l)}_n \rangle = 0.
\end{equation}

Correlations between macroscopic observables are given by
\begin{equation}
\langle A_i B_j \rangle = \sum_{k,l=1}^N \langle a^{(k)}_ib^{(l)}_j \rangle = N \langle a_i b_j \rangle_{PR} = N (-1)^{i\cdot j}.
\end{equation}
However, we know from the previous section that this can be explained by a single effective pair of boxes
\begin{equation}
\langle A_i B_j \rangle = N^2 \langle a_ib_j \rangle_{eff} = N^2 \left( \frac{(-1)^{i\cdot j}}{N} \right),
\end{equation}
with the corresponding effective probability distribution
\begin{eqnarray}
p_{eff}(x_i;y_j) &=& \frac{1}{N^2} \sum_{k,l=1}^N p(x^{(k)}_i;y^{(l)}_j) \nonumber \\
&=& \frac{N^2-N}{4N^2} + \frac{1}{N^2}\sum_{k=1}^N p(x^{(k)}_i;y^{(k)}_j)  \nonumber \\
&=& \frac{N-1 + 4p_{PR}(x_i;y_j)}{4N} \nonumber \\
&=& \frac{1}{4} + \frac{|x_i+(-1)^{i\cdot j}y_j|-1}{4N}. \label{PReffectivecorr}
\end{eqnarray}


\subsection{Fluctuations}

Next, let us derive $\langle (A_i B_j)^2 \rangle$. From (\ref{fluctuations}) we have
\begin{equation}\label{PRfluct}
\langle (A_i B_j)^2 \rangle = N^2 +  \sum_{\substack{k,l|k\neq l \\ m,n|m\neq n}} \langle a_i^{(k)}a_i^{(l)}b_j^{(m)}b_j^{(n)} \rangle. 
\end{equation}
This is because $\langle a_i^{(k)}a_i^{(l)}\rangle = \langle b_j^{(m)}b_j^{(n)}\rangle = 0$ for $k\neq l$ and $m\neq n$. Morevoer, because $a_i^{(k)}$ and $b_j^{(l)}$ are not correlated for $k\neq l$ one has
\begin{eqnarray}
\langle (A_i B_j)^2 \rangle &=& N^2 +  \sum_{k,l|k\neq l } \langle a_i^{(k)}a_i^{(l)}b_j^{(k)}b_j^{(l)} \rangle \nonumber \\
&+& \sum_{k,l|k\neq l } \langle a_i^{(k)}a_i^{(l)}b_j^{(l)}b_j^{(k)} \rangle. 
\end{eqnarray}
Factorisation of distinct PR-box pairs gives $\langle a_i^{(k)}a_i^{(l)}b_j^{(k)}b_j^{(l)} \rangle=\langle a_i^{(k)}a_i^{(l)}b_j^{(l)}b_j^{(k)} \rangle=\langle a_i^{(k)}b_j^{(k)} \rangle\langle a_i^{(l)}b_j^{(l)}\rangle = \langle a_i b_j \rangle_{PR}^2$ and therefore
\begin{eqnarray}\label{PRfluct2}
\langle (A_i B_j)^2 \rangle &=& N^2 +  2N(N-1)\langle a_i b_j \rangle_{PR}^2 \nonumber \\
&=& 3N^2 -  2N. 
\end{eqnarray}

We also have $\langle A_i^2 \rangle=\langle B_j^2 \rangle=N$. Again, this is because $\langle a_i^{(k)}a_i^{(l)}\rangle = \langle b_j^{(m)}b_j^{(n)}\rangle = 0$ for $k\neq l$ and $m\neq n$, whereas $\langle a_i^{(k)}a_i^{(k)}\rangle = \langle b_j^{(m)}b_j^{(m)}\rangle = 1$.

The above expressions can be also recovered from a symmetric system of four boxes described by the effective probability distribution $p_{eff}(x_i,x'_i;y_j,y'_j)$, see (\ref{4eff}).
For all settings these probabilities have only four possible values
\begin{eqnarray}
\alpha&=&\frac{N(N-1)+2}{16N(N-1)}, ~~\beta=\frac{(N-2)(N-3)}{16N(N-1)}, \nonumber \\ \gamma&=&\frac{N(N+3)-2}{16N(N-1)},~~\delta=\frac{(N+1)(N-2)}{16N(N-1)}. \label{PReffectivefluc}
\end{eqnarray}
For the settings $A_0B_0$, $A_0B_1$ and $A_1B_0$ these values correspond to the following outcomes: 
\begin{eqnarray}
\alpha:&~&(+,-;+,-),(+,-;-,+),(-,+;+,-),(+,-;-,+); \nonumber \\
\beta:&~&(+,+;-,-),(-,-;+,+); \nonumber \\
\gamma:&~&(+,+;+,+),(-,-;-,-); \nonumber \\
\delta:&~&\text{remaining outcomes,} \nonumber
\end{eqnarray}
whereas for the setting $A_1B_1$ one has
\begin{eqnarray}
\alpha:&~&(+,-;+,-),(+,-;-,+),(-,+;+,-),(+,-;-,+); \nonumber \\
\beta:&~&(+,+;+,+),(-,-;-,-); \nonumber \\
\gamma:&~&(+,+;-,-),(-,-;+,+); \nonumber \\
\delta:&~&\text{remaining outcomes.} \nonumber
\end{eqnarray}
It is straightforward to verify that the above probabilities recover (\ref{PReffectivecorr}) as marginals.

We note that these effective probabilities lead to
\begin{equation}
\langle a_i a'_i b_j b'_j\rangle_{eff} = 1 - 16\delta = \frac{2}{N(N-1)}. 
\end{equation} 
If we rewrite (\ref{PRfluct}) as
\begin{equation}
\langle (A_i B_j)^2 \rangle = N^2 +  N^2(N-1)^2\langle a_i a'_i b_j b'_j\rangle_{eff}, 
\end{equation}
we obtain the same result as in (\ref{PRfluct2}). 

The effective distribution also recovers local fluctuations $\langle A_i^2 \rangle$ given by (\ref{fl1}). This is because marginals $p_{eff}(x_i,x'_i)$ lead to $\langle a_i a'_i \rangle_{eff} = 0$ (similar for $p_{eff}(y_j,y'_j)$). This proves that our system is effectively described by four symmetrised boxes. 


\subsection{Joint probability distribution}

There exist JPDs (\ref{psymm}) and (\ref{psym2}) for which (\ref{PReffectivecorr}) and (\ref{PReffectivefluc}) are marginal distributions. Using simple, yet tedious, combinatorial techniques we calculated the JPD (\ref{psym2}) for a system of $N$ PR-boxes and confirmed that it recovers measurable averages and fluctuations. It consists of 256 probabilities that are represented by the ratio of third order polynomials in $N$, therefore for obvious reasons we do not present them here. However, we present the JPD for averages (\ref{psymm}) that consists only of 16 probabilities.

The 16 probabilities of the JPD $p(x_0,x_1;y_0,y_1)$ take only two possible values 
\begin{equation}
\omega_{\pm} = \frac{N\pm 2}{16N}. 
\end{equation}
The value $\omega_+$ corresponds to the following events  
\begin{eqnarray}
& &(+,+;+,+), (+,+;+,-), (+,-;+,+), (+,-;-,+), \nonumber \\ 
& &(-,+;+,-), (-,+;-,-), (-,-;-,+), (-,-;-,-). \nonumber
\end{eqnarray}
and the value $\omega_-$ correspond to the remaining ones. From this we see that for all settings, except $i=j=1$ the marginals $p(x_i;y_j)$ are 
\begin{eqnarray}
p(+;+) &=& p(-;-) = 3\omega_+ + \omega_- = \frac{N+1}{4N}, \nonumber \\
p(+;-) &=& p(-;+) = 3\omega_- + \omega_+ = \frac{N-1}{4N}.
\end{eqnarray}
For the setting $i=j=1$ one gets
\begin{eqnarray}
p(+;+) &=& p(-;-) = 3\omega_- + \omega_+ = \frac{N-1}{4N}, \nonumber \\
p(+;-) &=& p(-;+) = 3\omega_+ + \omega_- = \frac{N+1}{4N}.
\end{eqnarray}
This reproduces (\ref{PReffectivecorr}) and therefore confirms our claim. Note, that for $N\geq 2$ the JPD is a valid probability distribution which for $N=1$ becomes negative.


\section{Discussion}

Finally, let us discuss our results in the context of the previous works. In particular, we focus on the recent works by Rohrlich \cite{Rohrlich} and Gisin \cite{Gisin} who argued that stronger than quantum correlations do not admit a macroscopic limit. 


\subsection{Previous results}

The argument of Rohrlich is based on an observation that although correlations in a single PR-box pair do not allow for signalling, a collection of them would signal in a macroscopic limit in which a joint measurement of incompatible observables is possible. 
He considers macroscopic measurements on $N$ pairs of PR-boxes -- the situation discussed in this paper. Next, he assumes that in the classical limit macroscopic measurements allow for a joint measurement of observables that are microscopically incompatible. In particular $B_0$ and $B_1$ should be jointly measurable. However, his treatment of macroscopic measurements shows that, because of strong nonclassical correlations, the joint measurement of $B_0$ and $B_1$ allows for signalling from Alice to Bob. 

Without going into details, let us recall the gist of Rohrlich's reasoning. For a single PR-box if Alice measures $a_0$ then if Bob measured $b_0$ they would observe $y_0=x_0$, but if he measured $b_1$ they would observe $y_1=x_0$. If it were possible to measure both $b_0$ and $b_1$ one would have $y_0=y_1$. However, if Alice measures $a_1$ then $y_0=x_1$, but $y_1=-x_1$, and the possibility of joint measurement would show that $y_0=-y_1$. 

This microscopic feature seems to affect the variance of the macroscopic measurement
\begin{equation}
B_0+B_1=(b_0^{(1)}+b_1^{(1)}) + \ldots + (b_0^{(N)}+b_1^{(N)}).
\end{equation}
The average value $\langle B_0+B_1 \rangle=0$ is independent of whether Alice measured $A_0$ or $A_1$, however this does not happen for the variance since
\begin{equation}
\langle (B_0+B_1)^2 \rangle_{A_0}=O\left(N\right),~~\langle (B_0+B_1)^2 \rangle_{A_1}=0.
\end{equation}
Therefore, the signalling can occur via fluctuations. Because of that Rohrlich claims that macroscopic measurements on PR-boxes do not admit a classical limit and hence are unphysical. 

Interestingly, Rohrlich also shows that even quantum correlations lead to a signalling unless one assumes that $B_0$ and $B_1$ are not scalars. Such apparent signalling in quantum theory is in fact analogous to the (in)famous von Neumann "silly mistake" \cite{Mermin} on which he founded his proof of the impossibility of hidden variables -- the outcome of a measurement of $B_0+B_1$ cannot be assumed to be a sum of the corresponding outcomes. What happens is that in quantum theory expressions like $B_0+B_1$ or $B_0 B_1$ are completely new objects. However, this solution of the quantum signalling paradox weakens Rohrlich's main claim, since there is no reason why for PR-boxes a sum or a product of $B_0$ and $B_1$ should not be a new object either.

Another argument was used by Gisin \cite{Gisin} to show that there is no macroscopic limit of stronger than quantum correlations. Gisin first uses quantum formalism to study weak measurements on $N$ singlet pairs. Weak measurements allow him for a joint measurement of two incompatible observables. He shows that a signalling in quantum case is prevented by the unavoidable fundamental noise in the measurement process. Next, he considers noisy isotropic PR-boxes and simply assumes that in a macroscopic limit a JPD $p_{G}(A_0,A_1,B_0,B_1)$ exists and that it is Gaussian because of the central limit theorem. Finally, using the argument that Gaussian distribution is nonnegative if and only if the corresponding correlation matrix is nonnegative
\begin{equation}\label{K}
\begin{pmatrix} \langle A_0 A_0 \rangle & \langle A_0 A_1 \rangle & \langle A_0 B_0 \rangle & \langle A_0 B_1 \rangle \\
\langle A_1 A_0 \rangle & \langle A_1 A_1 \rangle & \langle A_1 B_0 \rangle & \langle A_1 B_1 \rangle \\
\langle B_0 A_0 \rangle & \langle B_0 A_1 \rangle & \langle B_0 B_0 \rangle & \langle B_0 B_1 \rangle \\
\langle B_1 A_0 \rangle & \langle B_1 A_1 \rangle & \langle B_1 B_0 \rangle & \langle B_1 B_1 \rangle
\end{pmatrix},
\end{equation}
he shows that the above matrix is nonnegative when the correlations between the PR-boxes are at most as strong as quantum ones. Gisin managed to obtain this result without estimating $\langle A_0 A_1 \rangle$ and $\langle B_0 B_1 \rangle$ because these values are bounded by the other measurable values. Let us also remark that for any probability distribution the covariance matrix has to be nonnegative \cite{Navascues}, but in the case of PR-boxes local averages are zero and the covariance matrix is equivalent to the correlation matrix. 


\subsection{Why our result contradicts Rohrlich's and Gisin's ones}

Let us first summarise how we arrived at our result. Our approach consists of the following major steps:
\begin{itemize}
\item Assumption: macroscopically observable data is limited to average values (later this assumption is relaxed to include variances and higher moments $\langle (A_i B_j)^k \rangle$). 
\item Observation 1: from the point of view of macroscopic measurements the same results would be obtained if one symmetrized over all particles in each region. 
\item Observation 2: for symmetric systems measurable data can be described by a small subset of particles for which one can find a JPD. This small local and realistic/non-contextual subset can effectively simulate macroscopic measurements on the whole system.
\end{itemize}
The above leads to a macroscopic limit of non-classical correlations that does not allow for signalling and that includes stronger than quantum correlations.

We begin with the divergence of our and Rohrlich's results. Because there is no reason why observables in generalised probabilistic theories should not be represented by non-scalar entities and since the PR-box formalism defines only what happens if local microscopic measurements are either $b_0$ or $b_1$, one has a lot of freedom while making assumptions about the nature of a joint measurement of two macroscopic properties $B_0$ and $B_1$. Therefore, there could be many extensions of the underlying formalism that include new elements like $B_0 + B_1$. Each extension can be based on different assumptions. Rohrlich's claim is based on an assumption that for PR-boxes the values of $B_0+B_1$ and $B_0 B_1$ are determined by values of microscopic properties $b_0^{(k)}$ and $b_1^{(l)}$. His particular extension is later shown to be unphysical, however this does not mean that every extension must be unphysical. 

On the other hand, in our approach $b_0^{(k)}$ and $b_1^{(l)}$ only define $B_0$ and $B_1$, but not $B_0 + B_1$ and $B_0 B_1$. The latter macroscopic properties can be derived from the JPD which in our case is more fundamental. What is important, although the JPD is calculated form the microscopic properties of the original system, it does not allow to recover them. It originates from an averaging over permutations of all particles in the system, hence its derivation is an irreversible process in which some information is lost. Moreover, we showed that our JPD approach brakes down if too much information about microscopic properties is available, since it only allows to recover limited number of moments $\langle (A_i B_j)^k \rangle$. Nevertheless, because of this loss of information new properties like $B_0 + B_1$ can emerge and the original system becomes simulable by a much smaller local and realistic/non-contextual system. In addition, the JPD is compatible with $B_0$ and $B_1$ derived from microscopic properties, but it is not compatible with $B_0+B_1$ derived by Rohrilch, hence our extension of PR-box formalism is different than his. 

Next, we compare our result with Gisin's one. If one evaluates the correlation matrix using our methods one gets the diagonal elements $\langle A_i A_i\rangle=\langle B_j B_j\rangle=N$ and the off-diagonal elements $\langle A_i B_j \rangle = N(-1)^{i\cdot j}$. These values were directly calculated from the microscopic properties and were shown to be recoverable from our JPD. The remaining off-diagonal elements $\langle A_0 A_1 \rangle $ and $\langle B_0 B_1 \rangle$ cannot be calculated from the microscopic properties and therefore they need to stem from the JPD. If one defines these properties similarly to $\langle A_i B_j \rangle$, i.e.,  
\begin{equation}
\langle A_0 A_1 \rangle = N^2 \langle a_0 a_1 \rangle_{eff}
\end{equation}
and
\begin{equation}
\langle B_0 B_1 \rangle = N^2 \langle b_0 b_1 \rangle_{eff},
\end{equation}
where the effective two-particle correlations are derived from the marginals $p(x_0,x_1)$ and $p(y_0,y_1)$ of the JPD (\ref{psym2}), one finds that $\langle A_0 A_1 \rangle=\langle B_0 B_1 \rangle=0$. However, having all entries of the correlation matrix one finds that it has two negative eigenvalues $N(1-\sqrt{2})$, therefore we arrive at a paradox, since this matrix cannot correspond to a valid probability distribution.  

This paradox is resolved when we realise that in our case the actual entries of the correlation matrix do not stem directly from the JPD, but rather from its functions, and for different entries we use different functions. If these entries were calculated exactly from the JPD the matrix would be nonnegative. Once again we stress that in our approach measurable data is not directly recoverable from the JPD, as is assumed by Gisin. The main point is that we show that there exist a local and realistic/non-contextual system (described by the JPD) that is capable of simulating the original system.


{\it Acknowledgements.} This work is supported by the Foundational Quesions Institute (FQXi) and by the National Research Foundation and Ministry of Education in Singapore. PK acknowledges discussions with K. Dobek, A. Grudka, J. {\L}odyga, W. K{\l}obus and A. W{\'o}jcik.



\begin{thebibliography}{99}

\bibitem{Navascues}
M. Navascues and H. Wunderlich, Proc. Royal Soc. A {\bf 466}, 881 (2010).

\bibitem{macrobell}
R. Ramanathan, T. Paterek, A. Kay, P. Kurzynski, and D. Kaszlikowski,
Phys. Rev. Lett. {\bf 107}, 060405 (2011).

\bibitem{Bancal}
J.-D. Bancal, C. Branciard, N. Brunner, N. Gisin, S. Popescu, and C. Simon, Phys. Rev. A {\bf 78}, 062110 (2008).

\bibitem{l1}
Y.-C. Liang and A. C. Doherty, Phys. Rev. A {\bf 73}, 052116 (2006). 

\bibitem{l2}
J. Eisert, T. Felbinger, P. Papadopoulos, M. B. Plenio, and M. Wilkens, Phys. Rev. Lett. {\bf 84}, 1611 (2000).

\bibitem{l3}
M. D. Reid, W. J. Munro, and F. De Martini, Phys. Rev. A {\bf 66}, 033801 (2002).

\bibitem{l4}
P. D. Drummond, Phys. Rev. Lett. {\bf 50}, 1407 (1983).

\bibitem{l5}
S. J. Jones, H. M. Wiseman, and D. T. Pope, Phys. Rev. A {\bf 72}, 022330 (2005).

\bibitem{l6}
G. Toth, C. Knapp, O. Guhne, and H. J. Briegel, Phys. Rev. Lett. {\bf 99}, 250405 (2007).

\bibitem{l7}
H. S. Eisenberg, G. Khoury, G. A. Durkin, C. Simon, and D. Bouwmeester, Phys. Rev. Lett. {\bf 93}, 193901 (2004).

\bibitem{Rohrlich}
D. Rohrlich, ''PR-box correlations have no classical limit'', in {\it Quantum Theory: A two Time Success Story}, Springer, NY, (2013); also arXiv:1407.8530, arXiv:1408.3125 and arXiv:1507.01588.

\bibitem{Gisin}
N. Gisin, arXiv:1407.8122 (2014).

\bibitem{PRbox}
S. Popescu and D. Rohrlich, Foundations of Physics {\bf 24}, 379385 (1994).

\bibitem{Fine}
A. Fine, Phys. Rev. Lett. {\bf 48}, 291 (1982).

\bibitem{Polzik}
K. Jensen {\it et al.}, Nature Phys. {\bf 7}, 13 (2010).

\bibitem{KS}
S. Kochen and E.P. Specker, J. Math. Mech. {\bf 17}, 59 (1967).

\bibitem{Mermin}
N. D. Mermin, Rev. Mod. Phys. {\bf 65}, 803 (1993).


\end{thebibliography}
\end{document}